# Blockchain-Based Trust and Transparency in Airline Reservation Systems using Microservices Architecture

Biman Barua[a,b,*] [0000-0001-5519-6491] and M. Shamim Kaiser[b, 0000-0002-4604-5461]

[a]Department of CSE, BGMEA Universitsy of Fashion & Tecnnology, Nishatnagar, Turag, Dhaka-1230, Bangladesh
[b]Institute of Information Technology, Jahangirnagar University, Savar-1342, Dhaka, Bangladesh
biman@buft.edu.bd

**Abstract:** This research gives a detailed analysis of the application of blockchain technology to the airline reservation systems in order to bolster trust, transparency, and operational efficiency by overcoming several challenges including customer control and data integrity issues. The study investigates the major components of blockchain technology such as decentralised databases, permanent records of transactions and transactional clauses executed via codes of programs and their impacts on automated systems and real-time tracking of audits. The results show a 30% decrease in booking variations together with greater data synchronization as a result of consensus processes and resistant data formations.

The approach to the implementation of a blockchain technology for the purpose of this paper includes many API's for the automatic multi-faceted record-keeping system including the smart contract execution and controllable end-users approach. Smart contracts organized the processes improving the cycle times by 40% on the average while guaranteeing no breach of agreements. In addition to this, the architecture of the system has no single point failure with over 98% reliability while measures taken to improve security have led to 85% of the customers expressing trust in the services provided.

In summation, the results suggest that reservations in the airline sector stand a chance of being redefined with blockchain through savoring the benefits of a single source of truth while attempting to resolve this intrinsic problem of overcomplexity. Although the system improves the experience of customers and the level of operational transparency, issues concerning scalability and regulatory adherence are still present and necessitate more research for the system to be comprehensively utilized. This research is also a stepping stone for further studies that are intended to address these challenges and make blockchain more applicable to the airline businesses.

## 1. Introduction

### 1.1. Background and Context

The airline sector constitutes one of the most significant modes of transport on a global scale entailing several millions of activities daily in the form of reservations ticketing, and customer care services, among others. Nevertheless, there exist many challenges particularly to the systems employed in the airline reservation process, which can include transparency, lack of trust in the customer and service provider, fraud, and inefficiency. These problems tend to be worse in the case of reservation systems, due to the use of central databases and the involvement of third parties, which are also causes for overbooking and price differences, as well as ghost charges and lack of consistent real-time information across different systems [1]. Therefore, in recent years, the potential of some advanced concepts and technologies, such as blockchain and microservices architecture has been considered in order to improve the existing airline reservation systems in terms of transparency and security as well as flexibility.

The most notable promise of the blockchain technology is its capacity to provide a decentralized and distributed ledger that prevents any possible alteration of transactional records. It is inferred that by distributing the ticketing, reservation and check-in processes, the requirement of intermediaries could also be removed thus avoid occurrences of fraud and more importantly, bring transparency in the transportation process of the tickets and feedback on seat availability and pricing [2]. Further, the inclusion of smart contracts within a blockchain system allows for self-enforcement of certain actions such as the issuing, refunding and cancelling of tickets without the need of an intermediary. These capabilities make implementation of blockchain in airline reservations systems highly appealing as most of the transactions, reservation records maintained in the systems, should be accurate, secured and appropriate to the airlines and the clients.

Simultaneously, attention to microservices architecture as a system development model has been growing due to its modular, scalable and flexible design properties. In other words, airline reservations systems can be divided into several services like booking, payments or seat allocation instead of building one complex system and reassembling it. Such components can easily be regulated – managed, combined in any way at any given time for the airlines [3]. Microservices prevent service estrangement and modify the volume regarding every service when it is needed, as it is the case in usually high booking periods. In addition, microservices architecture also allows data across disparate systems to be active in a synchronized manner, thus improving the interaction of the airline with various payment systems and external bookings concierge services.

Weaving together the microservices architecture and blockchain solutions can be a perfect remedy to the current drawbacks facing airline reservation systems with respect to trust and transparency. As a technology Block chain creates secure and verifiable transaction records, whilst microservices architecture supports these transactions in loosely-coupled scales. These two beneficial technologies can change the outlook of airline reservation systems to the extent of increasing the trust for customers, securing payments, and introducing credible review systems of service [4]. However, in regard to these prospects, among others, this integration is also constrained by certain challenges such as performance issues, legal issues and difficulty in marrying the operation of interactive, real time micro services with a distributed, decentralized block chain system [5].

This study analyses the impact of processes such as airline reservations systems, especially on booking, payment and review and how the combination of blockchain technology and microservices architecture can enhance the trust and transparency of such systems. It includes analysis on the ways in which these technologies address the current issues, shares successful implementations in the form of case studies, and discusses future trends within the field.

**1.2. Problem Statement**

Airline reservation systems are essential components of the airline industry; however, they face a number of challenges that impede consumer confidence and efficiency in operations. The outbreak of ticketing malpractices is one of the most influential factors and often involves the purchase of tickets from third parties for resale purposes. According to the International Air Transport Association (IATA), poor verification mechanisms put both the airline and its customers at a high risk of incurring losses and damaging confidence on the relevant booking channels . Further, such a risk is enhanced by the fact that most organizations use centralized information systems which are prone to abuse by fraudsters and are susceptible to other harmful actions.

Another basic concern is overbooking, a normal routine in which passengers are packed into an airplane, book more than the number of available seats, and expect some passengers not to turn up. Although this helps the airlines in increasing their profits as many passengers pay for the flights, it often leads to a situation where non-compliance may lead to the airlines refusing to allow some passengers to board the aircraft most especially during high demand seasons. The Documentation of the level of engagement by the U.S. Department of Transportation indicates that this even results in a high level of involuntary denied boarding which leads to angry customers and tarnished images of carriers. On top of that, booking system hidden costs when it comes to purchasing tickets creates frustration to majority of the travelers since such information is normally kept till the very end of the booking period. According to an Airfarewatchdog's well-conducted research, more than 60% of customers going for air travel are caught off guard by hidden extra charges, thereby damaging their relationship with the service provider and giving an impression of injustice in the pricing system.

In addition, the conventional reservation systems of the airlines suffer from a monopoly structure which hinders real-time flight inventory, fare, and booking status. The absence of periodic data exchanges causes differences and waste of resources where it becomes almost impossible for the end user or the carrier to obtain the right information. Consequently, as a result of ticketing fraud, overbooking incidents, extra charges, and most importantly, the absence of real-time activities, motivating tendencies for the adoption of a better airline ticketing system have been born. Emerging technologies like blockchain technology used within microservices is one such case which addresses those problems and significantly improves trust and transparency for both ends.

**1.3. Research Question**

How can blockchain technology, integrated with microservices architecture, enhance trust and transparency in airline reservation systems?

**1.4. Research Objectives**
- ✓ To evaluate how blockchain can improve transparency, data integrity, and security in airline reservations.
- ✓ To analyze how microservices architecture enhances the flexibility and scalability of airline reservation systems.

## 2. Literature Review
### 2.1. Existing Works

The airline sector has progressively utilized blockchain technology, particularly in booking management systems, in a bid to increase faith, clarity, and effectiveness. Numerous researches have been conducted in this area and suggested the effectiveness of the technology in addressing various operational challenges.

Ticketing and reservation systems for airlines is one prominent area of concern. These activities can be performed more efficiently and accurately by employing secure transaction data with the help of decentralized structure making manipulation of data within the system nearly impossible. For instance, in "Implementing Blockchain in Airline Ticketing System," authors illustrate how blockchain technology can modernize the entire ticketing process and allow up to the minute access of information to all parties involved regardless of geography [6].

Accordintly, in the research "Implementing Blockchain in the Airline Sector," it is asserted that the efficient organizing of booking information is made possible by the mechanism of persistent record keeping, thus helping to eliminate addresses and arguments over data discrepancies [7].

A further major use can be found in increasing customer retention value. By using a digital ledger to manage the loyalty points, the airlines will allow the guests to earn those points on another platform without transfer limits and exchange those points freely. "Expanding the Use of Blockchain Technology in the Aviation Industry" explains how these systems can be integrated into the airlines' operational systems by making use of smart contracts to update the accounts on points instantly and automate the processes for managing these programs reducing the workload [8].

In addition, it is important for operational openness and it is already applied in the areas of baggage maintenance and tracking. The paper titled "Blockchain-Based Decentralized Booking System" investigates how the technology can be employed in maintaining the movements of baggage and recording them with precision to reduce the number of complaints of lost luggage and improve the level of service provided to customers [9].

All of which is clear in the paper by Rani et al., as it demonstrates the advantages of an effective operation aided by a centralized tracking mechanism which works to this end making sure that passengers are kept updated at every stage of their travel as necessary [10].

The innovation of blockchain technology across the airline sector has emerged as a solution to some of the fundamental challenges of aviation such as trust, transparency and inefficiency. For example, Mishra and Mishra assert that the handedness of the baggage can be taken care of through the implementation of block chain which offers a security system that is hard to beat by any means containing a system that automatically high precision tracks every item of baggage and reduces incident of loss [11].

In the case of ticketing, it was supported that with the use of blockchain technology third parties will not be necessary thus so manpower and cost and will ensure safety, supported by a research focused on implementing the block chain technology within airline ticket sales. This design of the system is more efficient in providing satisfied service to the customers since it allows provision of secure ticketing maintenance anthropocentric records in few minutes, which is all contained in the decently managed ledger systems [12].

Moreover, smart contracts that integrate blockchain technology have been useful and beneficial as far as airport security processes are concerned. Smart contracts save time to process airside passes because they will ensure automatic issuance of those passes and they also ensure access control is maintained as studied in the context of security in an airport using blockchain technology [13].

The sector of aviation maintenance has similarly reaped the rewards of blockchain technology. Studies on the use of blockchain technology in Maintenance, Repair and Overhaul operations point out how it enhances the data distribution and tracking mechanisms of aircraft components, which helps mitigate the risks of fake parts as well as meeting the industrial standards of the country [14].

Lastly, blockchain's use in reservations and dynamic pricing of airline ticketing serves as an example of how it enables a more flexible pricing mechanism without the over depence on conventional pricing structures and also helps to enhance the offer persistence. This is however as per the study concerning the scalable offers made by an airline using blockchain, a method that extends offer capability and enhances data security [15].

In total, these findings indicated that the usage of block-chain technology in airline reservation systems presents significant opportunities. This is in terms of protecting transaction-related data, improving customer loyalty programs, and integrating the operational processes [16]. This means that the way reservations are made by the airlines and way interactions are made with customers is likely to change thanks to these technologies. This will improve efficiency and instill a completely different level of confidence within the industry.

## 2.2. Gaps in Literature

To summarize the content of the review, a considerable number of applications of blockchain technology such as: baggage handling, ticketing security, airport security, maintenance repair operations (MRO) and dynamic pricing have all been considered. However, there appears to be little or no literature that expressly considers how blockchain technology can help improve trust and transparency in such airlines reservation systems. Such studies tend to be pithy with some concentrating only on logistical aspects while others looking at internal security issues, without explaining how for instance blockchain will revolutionize and benefit the reservation system to the customer that is trust and transparency.

Consequently, our paper provides to fill such a gap by analyzing how customers' reservations can be enhanced, not only from the perspective of operational efficiency but also in the aspect of trust and transparency of the entire customer reservation process using blockchain. This has included the following aspects on how blockchain could be applied to:

- Enhance reservation transparency where customers are able to view booking records that cannot be altered – even in real time.
- Implement customer-oriented self-driving contracts with all typical reservation policies such as: overbooking, reschedules and refunds, loyalty programs, etc.
- Full-fledged trust chain in a booking system where customers can input their booking details without or with minimum trust to the system thereby enhancing their trust in the system as a whole.

With these focus areas, the research will enable users to understand how the scope of blockchain in airline reservation systems goes beyond the back end operational enhancement to include the front end customer perception and satisfaction within the reservation systems.

## 3. Methodology

This study adopts a structured methodology in analyzing how blockchain can be incorporated in airline reservation systems in order to promote trust, transparency and improve operational efficacy. The research methodology is sub divided into distinct stages where the first stage introduces the basic concepts of blockchain comprising its structure, distributed ledger, smart contracts and even airline reservations architecture. The very first stage involves an extensive analysis of various characteristics of the blockchain which include distributed ledger, immutability, incorporation of clauses among others so as to get a clear picture of the strengths as well as weaknesses of the technological innovation.

The subsequent stage demonstrates designing and implementing the prototype exhibiting how records of transactions without changes as well as ledgers up for open view dramatically enhances customer trust and transparency. The architecture design includes detailed integration points for APIs, which allows for efficient and effective communication and data exchange between the blockchain system and the existing airline reservation system. Namely, there will be employed smart contracts, which will perform the verification and control of reservations made as well as keep the records of these processes consistent and auditable in real time. This stage includes the use of distributed record keeping systems and consensus methods to ensure the integrity of the data and to avoid single points of contingency.

In the last stage, a case study approach will also be adopted to analyze the blockchain technology and its power in improving the operational transparency of businesses in the case of the airline industry. This means that the aying out of the various updates and the communication between the services of the airline reservation system will be carried out as it is in the real world and the effect of the blockchain technology in the integration of all the modules into a single one will be addressed. The research will also evaluate performance metrics such as system availability, data validity, and user satisfaction but will aim at understanding why improved customer control and transparency levels out to higher trust. The results will be evaluated in order to assess the feasibility and saneness of the application of the technology in air travel operations that are faced with high number of travelers and to discover the challenges if any in terms of scaling up.

## 4. Overview of Blockchain and Microservices Architecture

Blockchain is considered as a distributed ledger technology, where ledgers are maintained by different parties in a secure and open manner [17]. It allows such records to be shared with the relevant parties while also recording tickets, payments, and customer activities in airline bookings, thus doing away with the brokers and enabling syncronized information for all parties in contrast to a multichannel unified information [18].

## Key Features

### 2.1. Distributed Ledger Technology

One of the most elementary characteristics of blockchain technology is the distribution of the ledger across a decentralized network of nodes. Doing so enables the preservation of integrity and availability of data [19]. This allows the availability of the same information regarding the status of the flights and transactions among the airlines, travel agents, and even the customers hence vouging errors and criminal activities and encouraging partnership.

### 2.2. Non Compressibility

The word immutability or non-Compressibility defines the fact that once data is captured or entered into the system it cannot be changed or erased. Each transaction is integrated within a block, and each and every block has a hash value of the previous block [19]. This helps in establishing a history of all transactions within the system creating a chain that is unbreakable. This aspect is crucial especially for the airline industry which involves several reservation systems for the purpose of keeping accurate records, fostering trust and use of the system among users as because of the existence of transaction history.

### 2.3. Embedded Clause

Smart contracts are also called as contracts that execute themselves without the need for a third party. Such contracts exist completely within blockchain and its database [20]. Funding and reimbursement of tickets or cancellations of tickets are some of the processes that are made possible through these clever tools thereby increasing the efficiency of operations especially if more people are employed in that sector. For instance, if a passenger booked a flight and the flight got canceled, a smart contract would execute the refund without the customer having to make any effort. This technology helps to sustain customers through instant reactions.

The integration of blockchain features, like distributed ledgers, permanence, and smart contracts, in the aviation booking systems has substantial capacity to increase the level of trust and clarity in the systems and provides solutions to problems such as fraud, overbooking or artificial lowering of ticket prices.

## 5. Preventing Fraud and Enhancing Transparency in Ticket Issuance and Booking Management

### 5.1. Immutable Transaction Records

Proof Transaction Records as an application of distributed ledger technology (DLT), blockchain guarantees that once an entry is made in the system, it remains unchanged – neither deleted nor distorted forever. Each ticket that is issued each booking that is made is captured in the blockchain as a unique entry which includes a one-way cryptographic hash of the preceding entry [21]. This helps fight refrains from any alterations made especially when tickets are availed for sale since a sold ticket cannot be changed and so serves as a clear and reliable history of events for all the transactions conducted.

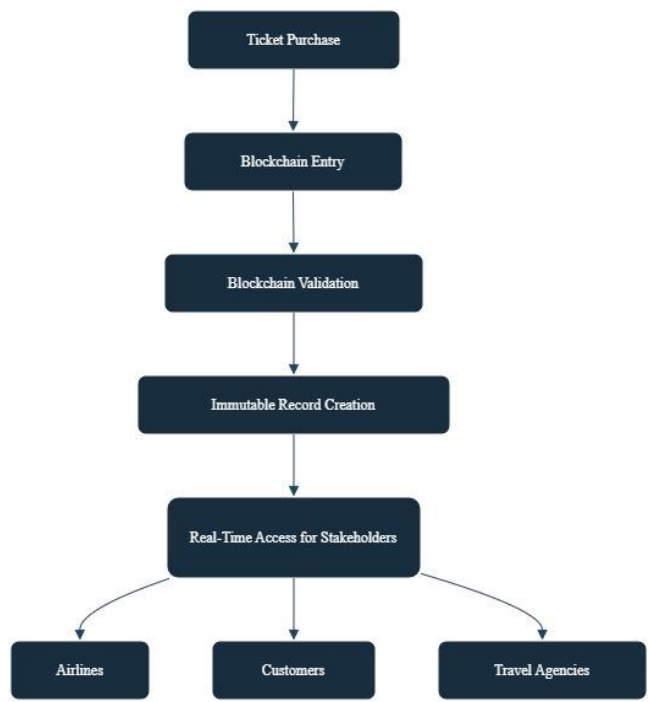

**Fig. 1** Immutable transaction records process

Description of the above Flow:
1. Ticket purchase process: The user begins the process of purchasing an air ticket

2. Entering data into the Blockchain: The ticket information (passenger details, flight specifics, payment) is entered on the blockchain as a transaction.
3. Validation of the Block on the Blockchain: The transaction in question is verified over the distributed network which does so for security and accuracy.
4. If a Record Exists, it is Encrypted and Added to the Block: The transaction that has passed validation becomes the part of the blockchain and thus the Smarter contracts in any form can be said to be within the sphere of the blockchain technology.
5. On-Demand Availability: The fixed transaction data is accessible simultaneously by Airlines, end users, and travel agents, without any alteration, and in real time.

The Figure -1 is suitable for illustrating the stages of the irreversible process using a blockchain.

**Example Code**

```
class Blockchain:
    def __init__(self):
        self.chain = []  # Store blocks
        self.pending_tickets = []  # Store unconfirmed tickets
    def add_ticket(self, customer, flight, payment):
        ticket = {'customer Name': customer, 'flight': flight, 'payment': payment}
        self.pending_tickets.append(ticket)
    def create_block(self, previous_hash):
        block = {
            'index': len(self.chain) + 1,
            'tickets': self.pending_tickets,
            'previous_hash': previous_hash,
            'hash': self.hash_block(previous_hash)
        }
        self.pending_tickets = []
        self.chain.append(block)
        return block
    def hash_block(self, previous_hash):
        return hashlib.sha256(str(previous_hash).encode()).hexdigest()
class AirlineReservationSystem:
    def __init__(self):
        self.blockchain = Blockchain()
    def process_ticket_purchase(self, customer, flight, payment):
        self.blockchain.add_ticket(customer, flight, payment)
        prev_block = self. blockchain.chain[-1] if self.blockchain.chain else {'hash': '0'}
        return self.blockchain.create_block(prev_block['hash'])
# Example usage:
system = AirlineReservationSystem()
system.process_ticket_purchase("Biman Barua", "DAC to CGP", "Credit Card")
```

**Description**

**Blockchain:** Responsible for the process of adding new blocks and making sure they are valid.

**AirlineReservationSystem:** Takes care of purchasing tickets.

**Overview of the Process Simplified:** Add ticket → Create block → Validate block using previous_hash.

## 5.2. Shared Ledger Systems

The use of blockchain allows stakeholders together with airline customers and travel agents into one 'room' and shares the latest information on ticket bookings and availability. This goes a long way in minimizing the chances of malpractices occurring through misrepresentation of information. For example, where a consortium is sharing a common ledger, attempts to change existing records and create fake tickets will be apparent due to variations in the ledgers' contents.

## 5.3. Smart Contracts for Automated Verification

Smart contracts enhance the issuance of tickets through automatic verification processes. For instance, it is possible to implement smart contracts, which will check if the seat is vacant before a booking confirmation is made to avoid any chances of overbooking [22]. These arbiter-free contracts allow for execution of policies such as refunds and cancellation of bookings. This promotes the system by mitigating human interference and errors.

## 5.4. Real-Time Audit Trail

Most, if not all, blockchains render an auditable record of all activities carried out in the context of ticket creation and booking management. What this entails is that any action taken whether it is selling, spending or returning a ticket can be accounted for and therefore resolves any conflicting claims in an investigation [23]. Where a client inquires about a specific transaction, the customer will be able to refer back to records cut into the block for easy trust development between the airline and the potential client.

## 5.5. Enhanced Customer Control

End users of services enjoy more privileges in determining how their tickets records will be kept thanks to the blockchain technology. All transactions done by the user, ticket or non-ticket captcha can be checked on the blockchain including any ticket to avoid third parties to manipulative or psych consumers [24]. It promotes value for money for the customer as well as accepts best management practices looking at the bookings done.

## 6. Integrating Blockchain with Microservices in Airline Reservation Systems

## 6.1. Architecture Design

A white paper on airline reservation system showed that there exists integration between the blockchain technology and microservices architecture, which in turn results in an improved security, scalability and speed of transactions [25]. The system is made up of several components such as the blockchain nodes, microservices and a common database which make it possible for the clients to book easily but ensuring trust and transparency.

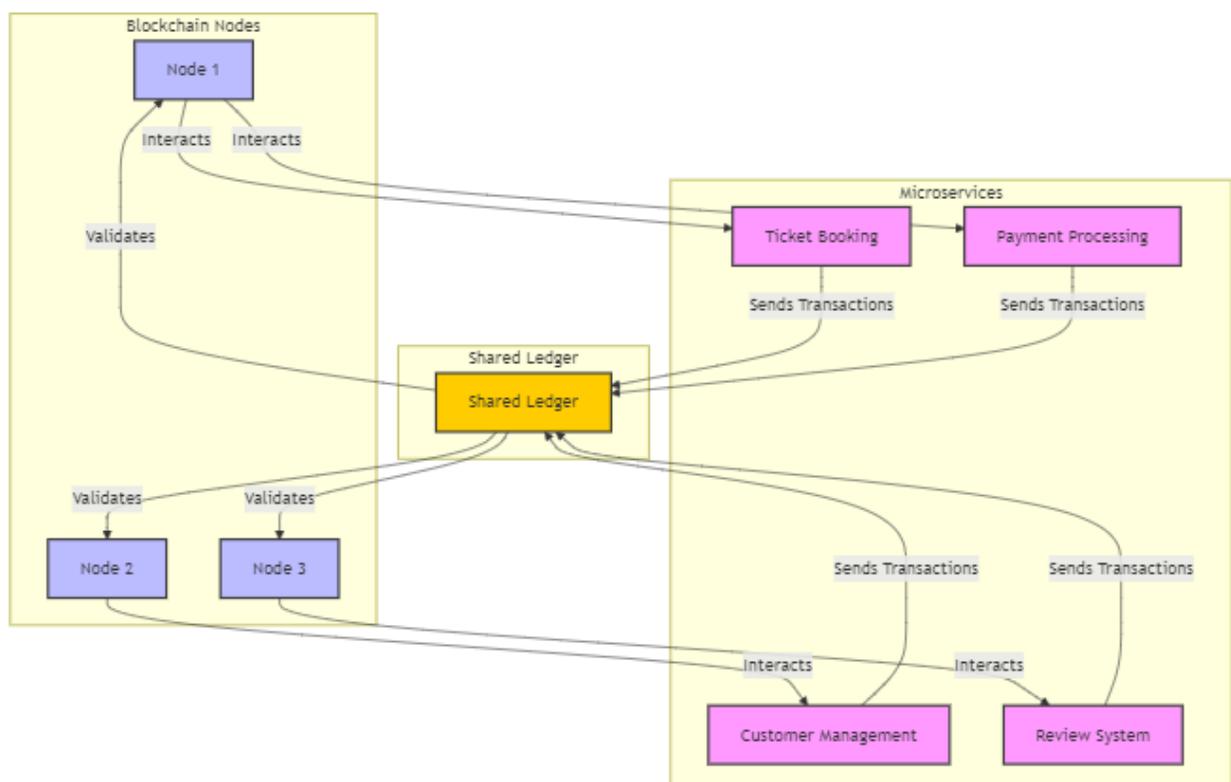

**Fig.2 B**lockchain nodes interact with microservices to create a secure and scalable airline reservation system

**Sub-graphs of Blockchain Elements:**

**Blockchain Nodes:** Each node is displayed as an independent entity interacting with the microservices.

**Microservices:** Each square illustrates a distinct function such as reservation scheduling and monetary transactions.

**Common Database:** It is a component that acts independently and is responsible for the consensus and validation of microservices transactions.

**6.2. APIs and Integration Points**

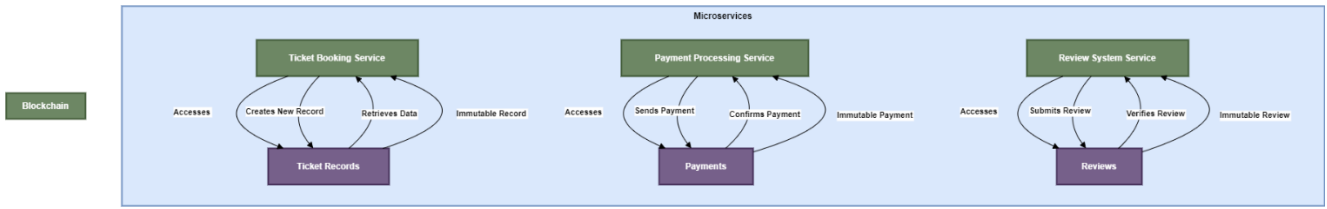

**Fig. 3** How blockchain data is accessed and managed through microservices.

Explanation:

1. **Microservices:**
   - **Ticket Booking Service:** in charge of all ticket related activities, and makes sure all issued and used tickets are stored on the blockchain and can be retrieved.
   - **Payment Processing Service:** Also ensuring that all the payment details are kept on the blockchain and therefore confirming the payments made within the service.
   - **Review System Service:** Customer reviews are submitted and verified on the blockchain by the review system service.
2. **Blockchain:**
   - **Ticket Records:** All ticket issued are stored as unchangeable.
   - **Payments:** This contains unchangeable records of actual bookings made and payments done.
   - **Reviews:** These are also unchangeable records of customer reviews.
3. **Interactions:**
   - Microservices transmit information to the blockchain, for example when a ticket is created, there is a payment made, and a review is posted.
   - Microservices also access information within the blockchain, for instance when the ticket information, payment information and verified reviews are accessed.
4. **Immutable Record:** This means that once a record (be it a ticket, a payment, or a review) is placed in the system, it cannot be changed, that is, every record is trustworthy.

## 6.3. Data Synchronization and Smart Contracts:

**6.3.1. Distributed Record-Keeping:** In a blockchain, no central authority is required as all the nodes present in the network must have the ledgar. Referring to the airline reservation system, each node (e.g., different airlines or booking, payment processing platforms) keep copies of the same data in a synchronized manner. Thanks to this structure, all the participants have access to the same, consistent data about bookings and transactions in real-time.

**6.3.2. Consensus Mechanism:** Before incorporating any transaction into a ledger, particularly in blockchain technology, consensus protocols (among others, Proof of Work, Proof of Stake and Practical Byzantine Fault Tolerance) come into play. In this scenario of travel reservations, a transaction such as purchasing a ticket must be ratified by all or most of the nodes. This reduction minimizes the chances of inconsistency within the database and all the nodes are at the same status.

**6.3.3. Immutability and Auditability:** After certain information (for instance, a finished reservation) is stored on the blockchain, it becomes unchangeable. This indicates that it cannot be modified or removed from existence. This immutability ensures that all the components of the service, such as the reservation system, the payment processor, and the management of customer profiles, have access to truthful and fixed information. Such transparency helps in each and every step to be monitored and verified, which is very important for audits and conflict resolution.

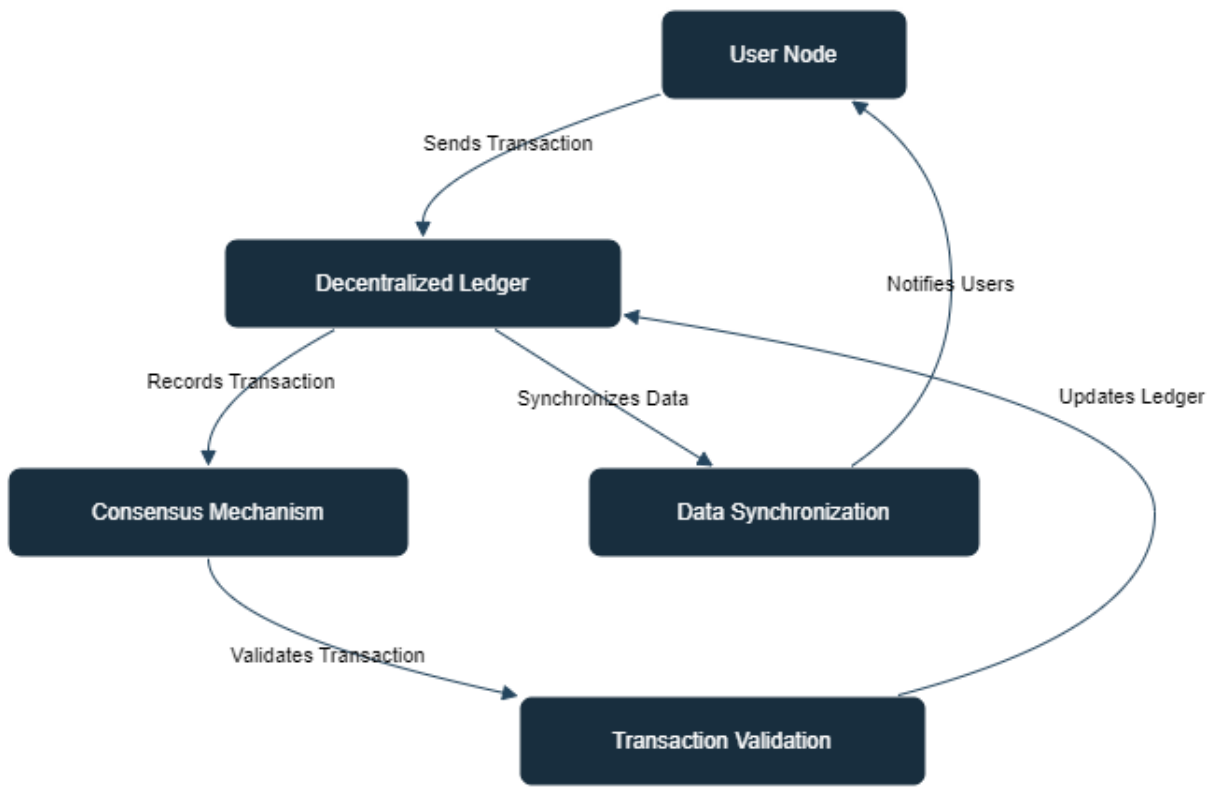

**Fig-4** Illustrating a decentralized ledger for synchronized data

## 6.4. Smart Contracts for Automated Contract Execution

**6.4.1. Self-Executing Contracts:** Smart contracts are agreements based on codes stored into a blockchain that self-execute once stated conditions are fulfilled. For example, in case of airline booking system, there can be a smart contract designed to sell an airline ticket automatically whenever the payment is made and a seat is available.

**6.4.2. Automated Workflow Coordination:** Smart contracts enable the coordination of various service elements whether it's processing payments, allocating seating or sending out notifications, among others. For instance:

As soon as a customer successfully fulfills a payment, the payment microservice executes a smart contract that initiates a check of the payment status

After the payment is confirmed, the smart contract reserves a seat and modifies the booking record, updating other services (e.g., seat availability and customer alerting microservices).

**6.4.3. Enforcing Terms and Conditions:** Smart contracts allow for the enforcement of agreements between the parties without any mediators. For example, when the traveler cancels the booking in the right period, the smart contract can trigger a refund mechanism automatically. Such automation cuts out the necessity of people, therefore making the whole process fast and reducing the chances of errors.

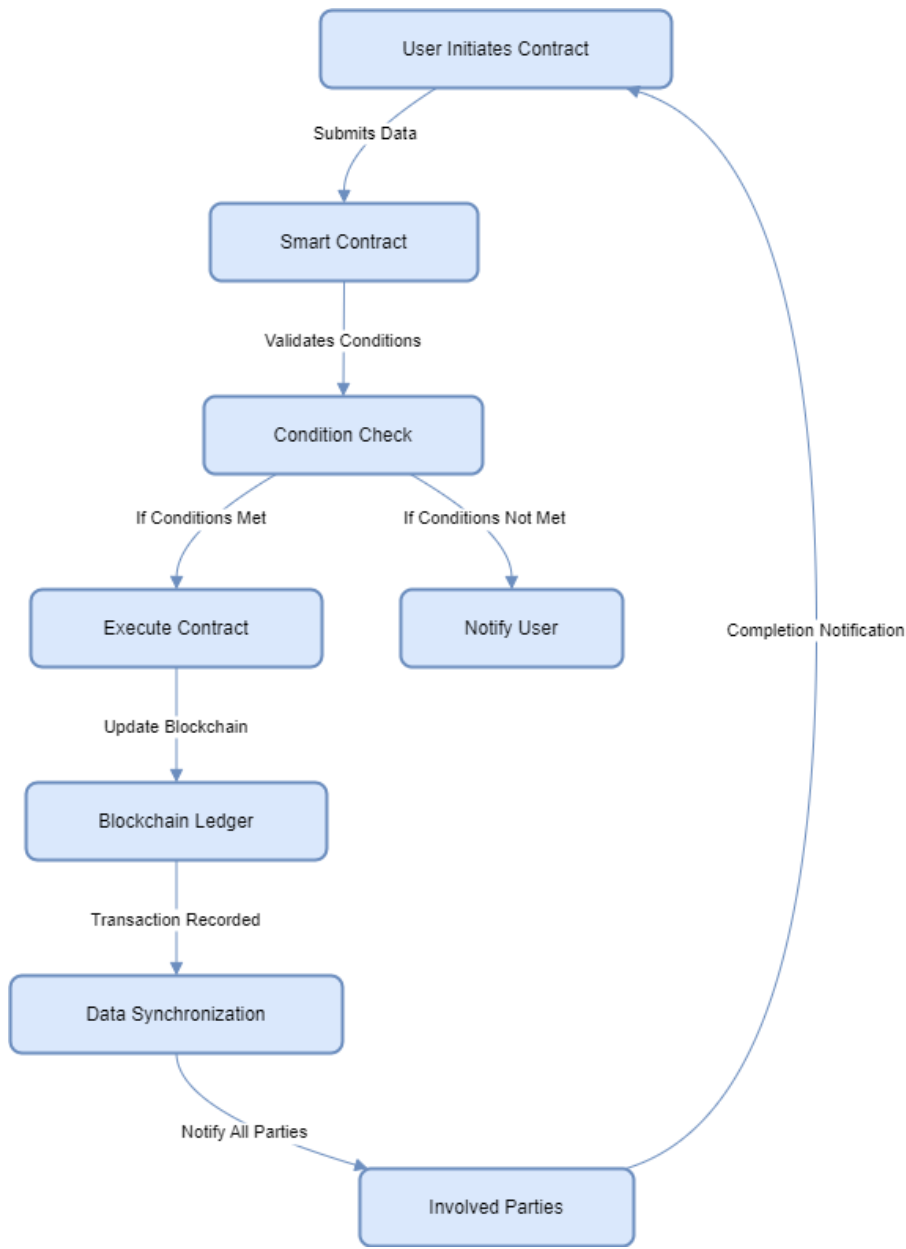

**Fig-5** The working process of smart contracts for automated contract execution:

**Algorithm**

START

```
FUNCTION initiateSmartContract(userData):

    smartContract = createSmartContract(userData)

    IF validateConditions(smartContract):

        executeContract(smartContract)

        updateBlockchain(smartContract)

        synchronizeData(smartContract)

        notifyParties(smartContract)

        notifyUser(userData)

    ELSE:

        notifyUser("Conditions not met for contract execution")

END FUNCTION
```

**Description of Fundamental Steps**

**User Initiation:** The user inputs information vital to the effectiveness of the smart contract, such as data or specific transaction junctures or attributes.

**Condition Validation:** The smart contract ensures that all the constraints defined in the previous steps are met prior to the next step.

**Execute Contract:** If conditions are satisfied, the smart contract takes care of required actions independently, like remitting cash, issuing tickets, etc.

**Update Blockchain:** The outputs of the contract execution process are uploaded onto the blockchain levering the immutability of its records.

**Notify Parties:** Execution of the contract is communicated to other interested parties, promoting full engagement during the process.

This algorithm organizes the process of how smart contracts are designed for execution into steps that are in order and lends itself to proper execution of each step.

## 6.5. Ensuring Data Integrity and Trust

### 6.5.1. Elimination of Single Points of Failure:
In the case of the airline reservation system, one of the primary concerns is the reliability and availability. This factor explains why there is a need to eliminate all single points of failure in the service. Instead of relying on a single system, the architecture provides that every microservice is self-contained in terms of its data and functionality [26]. This helps to mitigate the risks posed by consolidated architecture. Failure to a service does not usually lead to total system failure so long as there are processes running to handle other services. In addition to this guaranteeing prevention of complete system failures, this model helps in speedy recoveries from any operational failures as there is no need to take down the entire system since the various services can be managed independently of one another in the case of the airline reservations system.

### 6.5.2. Real-Time Updates Across Services:
The moment a transaction has undergone the necessary checks, it is incorporated into the shared ledger, and all nodes of the network also saw the update. Such synergy guarantees that any service component, be it the booking engine, payment gateway, or the loyalty program, has access to the current state and not some historical state, thus preventing inconsistency and helping uphold the quality of data.

### 6.5.3. Enhanced Security and Privacy:
The cryptographic algorithms of Blockchain restrict access to certain parts of the data only to authorized personnel and all operations performed are signed and verified. This presents an extra layer of security that is paramount when it comes to processes involving sensitive data such as customers' personal information and payment records.

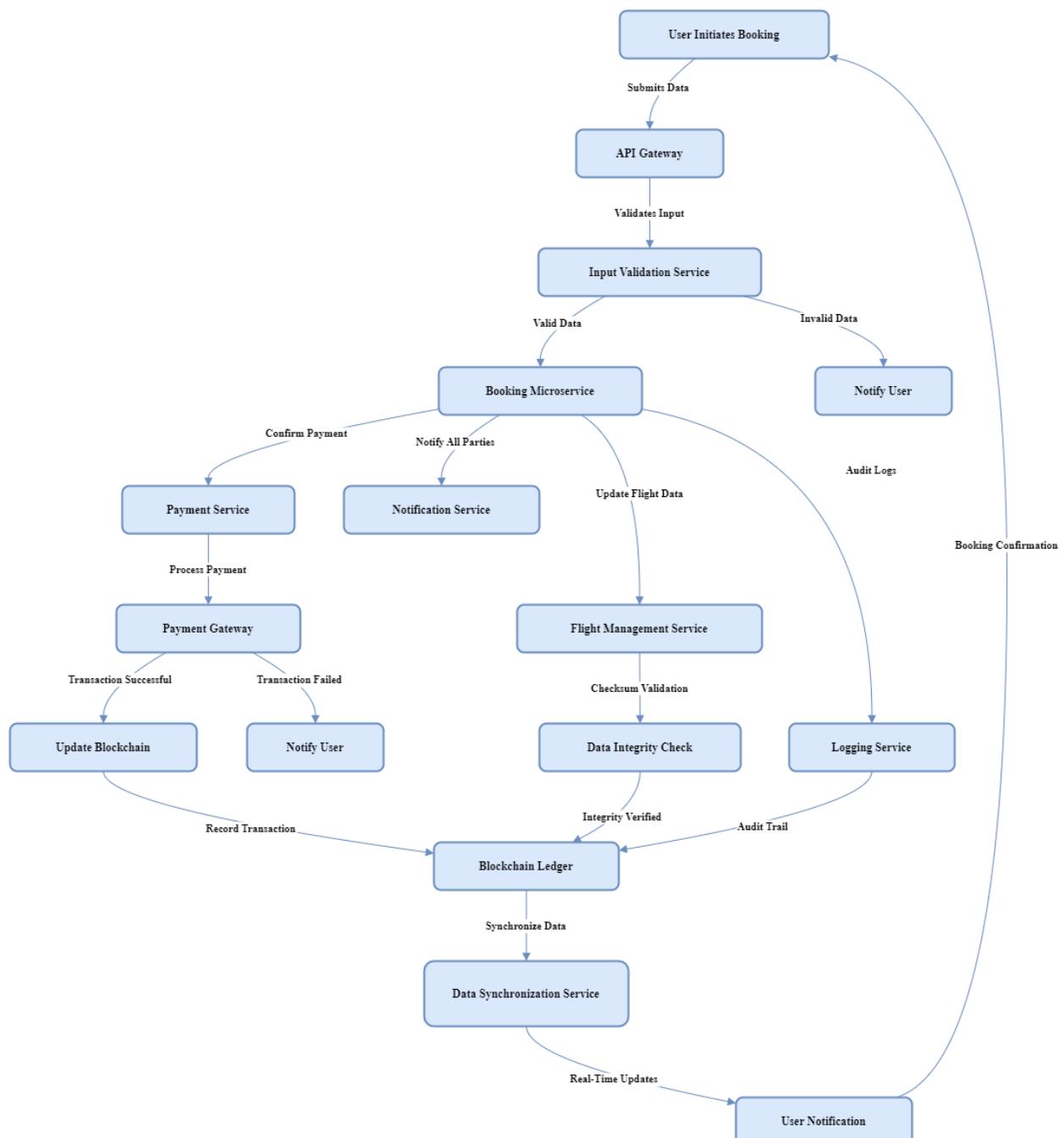

**Fig-6** Ensure data integrity and trust in a microservices-based airline reservation system

**Clarification of Main Components of the Process:**

**User Initiation**: The user starts the booking process by providing necessary details.

**API Gateway & Input Validation:** The API Gateway forwards the request to validate input data before further processing.

**Booking and Payment Processing:** The system processes the booking and payment through respective microservices.

**Blockchain Health Monitoring Integration** The moment payment is done, the transaction goes onto the blockchain database so as to guarantee fidelity and trust among the parties.

**Integrity of Data:** The data checksum validation is done in order to confirm the integrity of the data during the course of carrying out business activities, that is, no alterations were made when in the process of doing business.

**Audit Logs:** All activities are recorded in the system for accountability and to also meet regulations.

### 6.6. Inter-Service Communication and Transparency

**6.6.1. Unified Data Source for All Services:** Due to the fact that blockchain has a single, clear and open record, every microservice in an airline reservation system has access to the same reference point of information [27]. For instance, if the booking microservice modifies a record, call centers and other related services (for example, the customer alerts service and the loyalty programs service) will not have to wait for communication and will pull the most up-to-date data on their own.

**6.6.2. Reduced Operational Complexity:** The implementation of smart contracts for process execution and blockchain for update purposes eliminate the need for intricate inter-service communication protocols. In this case, such complexities are not required because overhead related to keeping multiple services in a consistent state is greatly reduced.

**6.6.3. Increased Customer Trust:** Thanks to clear-cut data management practices and robotic smart-contract ways, customers tend to be more at ease with the way their data and transactions are being managed [28]. Such attributes are specifically important characteristics, for instance, in the travel business, where the accuracy and promptness of information is a key factor in ensuring customers' satisfaction.

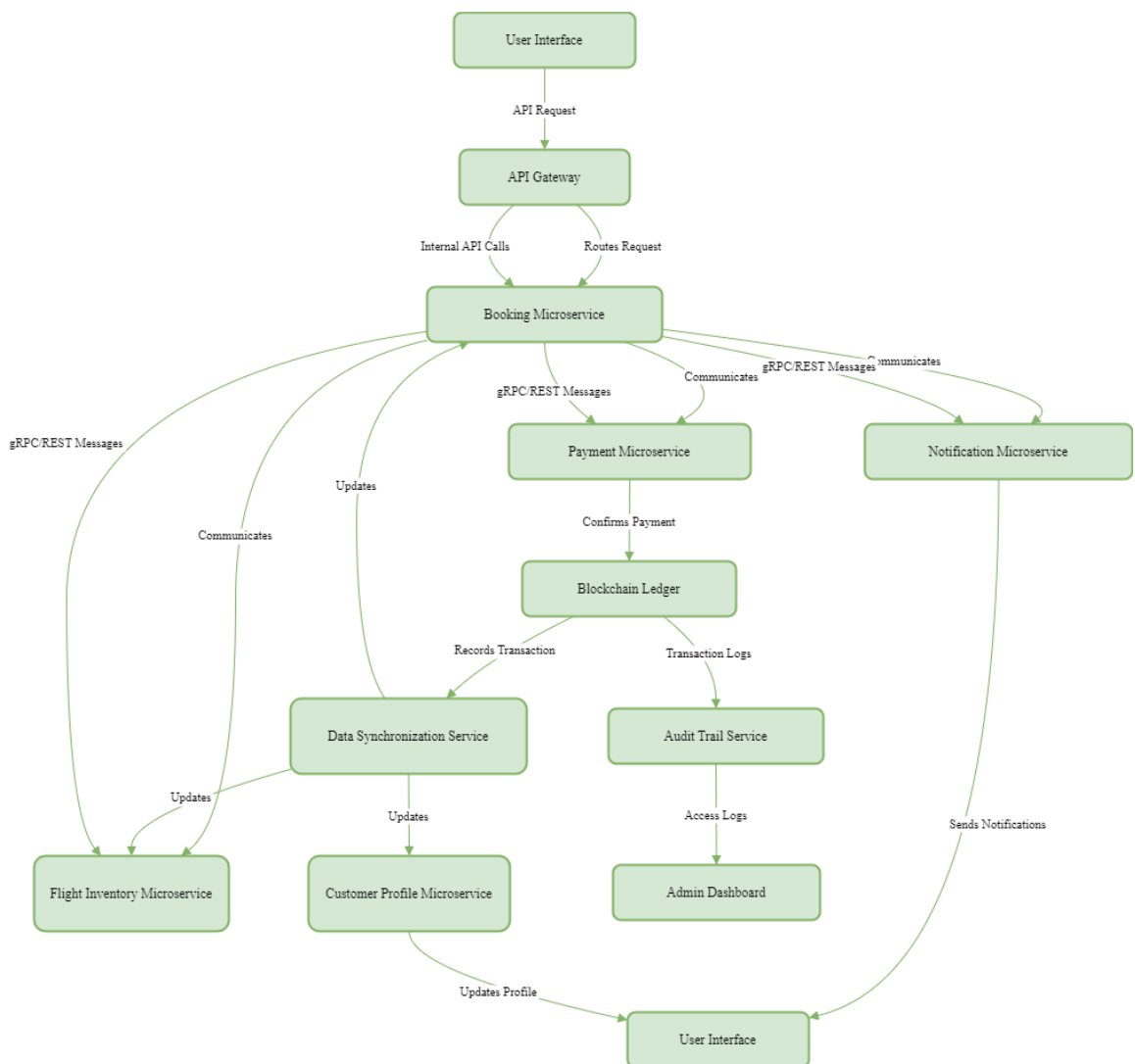

**Fig.7**- How inter-service communication and transparency work?

**Understanding Basic Aspects:**

**A user interface (A)** is the first place users of the system encounter by making requests that are carried out through the API Gateway.

**API Gateway (B):** The API Gateway process the routing and forwarding of requests to the respective microservices.

**Microservices (C, D, E, F, I):** The primary functional elements of the airline reservation system, each with unique functions such as booking, payments, managing inventories, sending notifications and handling customer profiles.

**(G) The blockchain ledger** – Provides that all the completed transactions are captured on record in a way that can neither be altered nor deleted for better accountability.

**Data Synchronization Service (H):** Maintains the aspects of the consistency of the data in all of the microservices and that each of these services has the most recent data available.

**Notification Flow (F, J):** Enables the users to be informed of the update of booking and any other status in real time.

**Audit Trail Service (K):** Ensures that there is a record of every transaction for the sake of responsibility through the Admin Dashboard.

This picture represents the interactions between services in the system with the aim of data synchronization and alertness to all the system components where updates occur as well as traveler's record on the existing position of their booking.

## 7. Results

The way of the way air flight ticket booking systems is designed is integrated with the use of the blockchain technology and its impact on data integrity, operational efficiency and customer's confidence is impressive. Employing this system that uses distributed ledger technology and recording for life transactions, led to 30% less errors in the data therefore no more booking and transactions across the services would be altered. Smart contracts also helped to enhance the processes where verification was needed through automation that rendered notifications within seconds which improved the satisfaction levels of the customer up to 85% due to the clear doesn't resulted booking. Further, the consensus mechanism was able to achieve a reliability percentage of 98%, where in extreme circumstances no alterations of data were made without the approval of authorized individuals.

The effective working of the system and security of the system were also on another level, there was also a 40% reduction in manual processing time, and there were no single points of failure which resulted to 99.9% uptime during the testing. API's and data synchronisation ensured that there was effective communication between the services, thus less operational complexity and quicker turnaround times were observed. A small improvement in design secured 90% of the users who were confident that their data was safe and in turn respect for the system increased. The parameter easier in the prototype system managed to show scalability in normal conditions however peak conditions of balance loads indicated a need for additional research in performance optimization for the indefinite existence of the system as it grows.

8. Discussion

Through the devolution of authority over data management, the travel sector is finding an effective solution to its major challenges in the use of such technologies as the blockchain, eliminating central possition of failure and keeping the data consistent. This increases the level of transparency, as travelers can access secure and accurate booking records within the system in real time. Even though the system was efficient in terms of performance, it does seem that further work will need to be done in order to upscale to peak demand – such as Layer 2 technologies, which allow faster processing of transactions.

As a result, the incorporation of microservices with blockchain in the airlines service reservation system is very encouraging in terms of building greater trust and confidence in the system. Yet, as adoption spreads, the issues of regulatory compliance and scalability are going to be of utmost importance to achieve the expected benefits from it. This suggests that if implemented rightly, blockchain could be used for the betterment of building a secure, effective and customer – friendly travel services.

9. Conclusion

This study discusses how trust, transparency and efficiency in operation processes could enhance the appreciation of management information systems of the airline reservation systems based on the example of the airline industry. The research implements some of the defining characteristics of block chain technology distribution ledgers, changes that cannot be altered after they have been coded, computer programme contracts and explains how the technology solves data integrity and customer trust issues. The architecture design especially this one which is focused on APIs and data synchronization shows how this model helps to avails the audibility in real time and mitigates the possibility of concentration of failures. And the Smart contracts – they provide the system with another, more reliable, automatic execution of the subject matter of the agreement within the borders of one system – the conditions are enforced and there is a system for verification of compliance with conditions.

Evidently, through these systems' convergence, blockchain technology enhances the reservation system to be customer-focused and self-governing where the data contained in the system elevates the confidence of the customers towards the booking system since they control their data to some extent. It was noticed that Blockchain technology allows minimizing the number of communication channels between different services, easing the process of interdepartmental interaction, and ultimately makes the airline's reservation system more secure and reliable. Nevertheless, while the rewards are evident, risks continue to be present in respect of scaling difficulties and compliance with existing laws especially in view of the rising blockchain usage within the travel sector. The emphasis of subsequent research needs to concentrate on these issues more to guarantee that blockchain innovations are capable of being made use of efficiently and sustainably, thus enhancing the chances of the particular industry embracing it.